\documentclass[11pt,a4paper]{article}
\pdfoutput=1 
\usepackage{jheppub}
\usepackage{physics}
\usepackage{amsmath}
\usepackage{color}
\usepackage{cases}
\usepackage{tensor}

\newcommand{\floor}[1]{\ensuremath \left\lfloor #1 \right\rfloor}
\newcommand*{\curv}{\mathcal{K}}
\newcommand*{\dual}{\vb*{\zeta}}
\DeclareMathOperator{\AdS}{AdS}

\newcommand*{\Eperp}{\mathcal{E}_{\perp}}
\newcommand*{\xieff}{\xi_{\text{eff}}}

\newcommand*{\MPl}{M^{\text{Pl}}}
\newcommand*{\xiperp}{\xi_{\perp}}

\title{Emergent Gravity in Spaces of Constant Curvature}
\author[a,1]{Orlando Alvarez%
\note{This work was supported in part by the National Science Foundation under Grant PHY-1212337.}}
\author[a]{and Matthew Haddad}

\affiliation[a]{Department of Physics, University of Miami, 1320 Campo Sano Ave, Coral Gables, FL 33146, USA}

\emailAdd{oalvarez@miami.edu}
\emailAdd{m.haddad@miami.edu}

\abstract{In physical theories where the energy (action) is localized near a submanifold of a constant curvature space, there is a universal expression for the energy (or the action).  We derive a multipole expansion for the energy that has a finite number of terms, and depends on intrinsic geometric invariants of the submanifold and extrinsic invariants of the embedding of the submanifold.  This is the second of a pair of articles in which we try to develop a theory of emergent gravity arising from the embedding of a submanifold into an ambient space equipped with a quantum field theory.  Our theoretical method requires a generalization of a formula due to by Hermann Weyl.  While the first paper discussed the framework in Euclidean (Minkowski) space, here we discuss how this framework generalizes to spaces of constant sectional curvature.  We focus primarily on anti de Sitter space.  We then discuss how such a theory can give rise to a cosmological constant and Planck mass that are within reasonable bounds of the experimental values.
}

\begin{document}
	\maketitle
	
	\section{Introduction}
 
There is a general phenomenon that occurs when the action (energy) of a Minkowskian (Euclidean) field theory is localized near an embedded submanifold. The effective action that determines the dynamics of the embedded submanifold  has a universal form that includes, among its terms, the generalizations of general relativity by Lovelock-type Lagrangians. Even though no gravitation was assumed \emph{a priori}, what emerges is a gravity-like theory that describes the dynamics of the submanifold. This article is an extension of a recent work by one of us~\cite{alvarez:emergentflat} where this mechanism was discussed for the case of embeddings of submanifolds in Minkowski (Euclidean) space. Here we extend the results of that paper to  embedding the submanifold in a space of constant curvature. We often specialize our  discussion to the case of  anti de Sitter space $\AdS_{n}$ because of its central role in  theoretical physics.    $\AdS_{n}$ is a Lorentzian manifold with constant negative sectional curvature $k$. We define its radius of curvature to be $\rho = \lvert k \rvert^{-1/2}$.

The main technical result of this paper is the development of a multipole expansion for the energy (action) analogous to the one developed in the companion article. The main physical result is a new mechanism for an emergent gravity-like theory where the constant curvature of the manifold introduces an additional length scale not present in the flat-space embedding models discussed in the companion article. In this article, we study embedding in $\AdS_{n}$ with curvature compatible with the observed bounds on that of our 4-dimensional universe. The emergent theory of gravity has a 4D cosmological constant $\Lambda_{4}$ that is in agreement with the experimentally observed value\footnote{All computations in this article are classical.}. The value of $\Lambda_{4}$  is largely independent of value of the 4D Planck mass $\MPl_{4}$. Starting with essentially a massless higher-dimensional theory with a very low energy scale (on the order of $10^{-22}~\text{GeV}$), we find that we can generate a 4D Planck mass $\MPl_{4}\sim 10^{19}~\text{GeV}$. Such a small energy scale could arise in a conformal $n$-dimensional field theory where the isometry group of $\AdS_{n}$ is broken very softly.

There are two central topics in this paper. The first is the emergence of a gravity-like theory. The mechanism and related issues are very similar to the flat-space case discussed in  \cite{alvarez:emergentflat}, and we will not repeat the discussion here. The second topic concerns the values of the cosmological constant $\Lambda_{4}$ and the Planck mass $\MPl_{4}$ in four dimensions. The literature on these subjects is vast and we will concentrate on the relationship of our work to that discussed in
\cite{ArkaniHamed:1998rs,ArkaniHamed:1998nn,Randall:1999vf,Randall:1999ee,Dvali:2000hr}. In references~\cite{ArkaniHamed:1998rs,ArkaniHamed:1998nn}, Arkani-Hamed, Dimopoulos and Dvali (AHDD) consider a Kaluza-Klein compactification-type scenario in a theory of higher-dimensional gravity with a TeV mass scale for the gravitational force and a compactification radius of roughly $1$~mm to induce the weak gravitational force seen in four dimensions at distance scales larger than $1$~mm with strength given by $\MPl_{4} \sim 10^{19}~\text{GeV}$. In this scenario, the TeV scale is motivated by the electroweak interactions and the 1~mm compactification scale is chosen to get the correct $\MPl_{4}$. The scenario of Randall and Sundrum (RS) \cite{Randall:1999vf,Randall:1999ee} uses a piece of $\AdS_{5}$ that has a boundary consisting of two $3$-branes. The world we inhabit is one of the two boundary pieces. As one moves from one boundary to the other, the induced metric changes exponentially. Randall and Sundrum use this to generate an exponential hierarchy between the TeV scale and the Planck scale. Their ``compactification scale'' $r_{c}$ is of the order of the Planck length. The four-dimensional gravity in the Randall-Sundrum model is induced by the higher, five-dimensional gravity theory in $\AdS_{5}$. In the scenario presented by Dvali, Gabadadze and Porrati (DGP) \cite{Dvali:2000hr}, the gravitational Lagrangian consists of two parts. A four-dimensional piece that lives in a $3$-brane and a five-dimensional piece in an infinite bulk. These authors show that they can reproduce the observed $1/r$ gravitational potential at the appropriate distance scale in the $3$-brane.

The model studied here has aspects of the three scenarios reviewed in the previous paragraph but with some important differences. The first difference is that we assume that there is  no fundamental theory of gravity in the embedding manifold $\AdS_{n}$. There is an embedded $q$-dimensional submanifold  $\Sigma^{q} \hookrightarrow \AdS_{n}$, a $p=(q-1)$-brane, where the action (energy) is localized near the submanifold\footnote{We use the notation $n=q+l$.}. Just as shown in \cite{alvarez:emergentflat} we find that the dynamics of $\Sigma^{q}$ are determined by a Lovelock-type theory of emergent gravity. We develop a multipole expansion for the action (energy) density that allows us to systematically compute the induced gravitational parameters generalizing the results of \cite{Forster:1974ga,Maeda:1987pd,Gregory:1988qv,Gregory:1990pm}. To explore the effects of the curvature of $\AdS_{n}$, we assume that the masses of the fundamental particles are very small, so that the associated Compton wavelength is much larger than the radius of curvature of $\AdS_{n}$. This radius of curvature acts as an effective length cutoff even though our $\AdS_{n}$ is infinitely large. In this way we get induced gravitational parameters on $\Sigma^{q}$ that are reminiscent of the Kaluza-Klein-type theories \cite{ArkaniHamed:1998rs,ArkaniHamed:1998nn} with the radius of curvature $\rho$ playing the role of the compactification radius\footnote{The localization of the energy near the submanifold $\Sigma^{q}$ in this model is over a cosmological distance.}. Since there is no higher-dimensional gravity, the higher-dimensional energy scale is set by the higher-dimensional field theory and not by the higher-dimensional gravitational constant. We have a mechanism in which the effective compactification radius arises naturally from the radius of curvature of spacetime which we take to be compatible with the experimental bounds. These give a length that is approximately the observed radius of the visible universe $\rho\sim 10^{10}~\text{ly} \sim 10^{26}~\text{m}$. We do not have $n$-dimensional gravity, therefore we do not have to worry about the crossover from the mathematical form of the gravitational potential in the bulk $\AdS_{n}$ to the mathematical form of the gravitational potential in the $\Sigma^{q}$ worldbrane. In flat space this would be the crossover behavior from $1/r^{n-3}$ in the bulk to $1/r^{q-3}$ on the worldbrane. In our scenario, we do not need the DGP mechanism to solve the crossover issue because we do not have $n$-dimensional gravity. We differ from the original RS scenario because their radius of curvature for $\AdS_{5}$  is of order the Planck length, $10^{-35}~\text{m}$, and their ``compactification radius'' $r_{c}$ is an order of magnitude larger. The RS universe is a very narrow slice of $\AdS_{5}$. In our approach, the radius of curvature of $\AdS_{n}$ is of cosmological size, and the $n$-dimensional energy scale is minuscule as we will discuss later. 

Our discussion is often guided by the properties of static topological defects in $\AdS_{n}$. The presence of curvature changes the asymptotic behavior of the fields and this is discussed in a forthcoming paper~\cite{Alv:2016c}.  Here we only use the general assumption that the energy density decays exponentially in directions orthogonal to the submanifold $\Sigma^{q}$.   What we discover is that, at large distances from the submanifold, the exponentially increasing volume of a tubular region surrounding the submanifold can potentially compensate for the exponentially decreasing energy density and lead to physical manifestations where the curvature of $\AdS_{n}$ determines the lower dimensional cosmological constant and Planck mass. An interesting result of our scenario is that the four-dimensional cosmological constant is given by $\Lambda_{4}\sim 1/\rho^{2}$ which is consistent with the currently measured experimental value.

	\section{Review of results in flat space}
	It has been shown~\cite{weyl:tubes} that for a $q$-dimensional submanifold $\Sigma^{q}$ embedded in $n$-dimensional Euclidean space that the volume element for some tubular region near the submanifold can be expressed as:
	\begin{equation}
	\dd[n]{x} = \det(I + \vb*{\nu}\vdot\vb{K})\, \vb*{\zeta}_\Sigma \wedge \dd{\nu^{q+1}} \wedge \dd{\nu^{q+2}} \wedge \cdots \wedge \dd{\nu^n}
	\end{equation}
	where $\vb*{\zeta}_\Sigma$ is the induced volume element on the submanifold and $\nu^{i}$ are Cartesian coordinates on the fibers of the normal bundle of the submanifold, see \cite{alvarez:emergentflat}	for the notation and a modern derivation.
	
	As a brief recap of the flat-space case, consider the vortex solution to the equations of motion in the Abelian Higgs model in ordinary $n$-dimensional Minkowski space $\mathbb{M}^n$. The vortex is described by the locus $\qty{x\in\mathbb{M}^n \mid \phi(x) = 0}$, where $\phi$ is the complex scalar field. The vortex arises from a spontaneously-broken $U(1)$ symmetry and imposes a topological quantization condition. This vortex is an $(n-2)$-dimensional submanifold  $\Sigma$ of the ambient space.
	
	In the case of a spherically symmetric vortex solution, the energy density is exactly of the right form to apply the Weyl volume formula. One can then form an expansion of this energy in a finite number of radial monopole moments. In this particular example, the moments are constants. Denote by $T_{q}$ the $p$-brane tension, and by $G_q$ the $q$-dimensional Newtonian gravitational constant. Calculating the first two monopole moments gives the vacuum and matter energy contributions, the ratio is $T_{q} G_q \sim 1/\xiperp^2$, where $\xiperp$ is a correlation length of the theory~\cite{alvarez:emergentflat}.
	
	\section{Weyl's volume element formula in a space of constant curvature}
	
In Hermann Weyl's	original article~\cite{weyl:tubes}, he extends his tube volume results to the case of embedding the submanifold in a space of constant positive curvature. Here we present a different derivation that is valid for \emph{any} constant curvature space by constructing in a space of constant curvature the analog of the Darboux frame extension of~\cite{alvarez:emergentflat}. The only constraint required for the validity of the derivation presented here is that the normal directions to the submanifold $\Sigma^{q}$ have Euclidean signature. Consider a $q$-dimensional submanifold $\Sigma^q$ embedded into an $n$-dimensional space of constant curvature $M^n$. We will use an adaptation of Cartan's approach to Riemannian normal coordinates~\cite{cartan:riemannian} as described in \cite{alvarez:schwarzschild} adapted to our situation\footnote{There is a discussion of the use of Fermi normal coordinates in tube studies in Gray's book~\cite{Gray:tubes} but it is very different from  ours.}. Cartan's method constructs an orthonormal coframe and this is the analog of the Darboux coframing extension in the flat-space case. We use parallel transport to extend the Darboux frame away from the submanifold.

Choose a point $\sigma \in \Sigma$ and consider a geodesic $\gamma(\lambda)$ beginning at $\sigma$ with initial velocity $\vb{u}$ orthogonal to $\Sigma$. Note the initial conditions for the geodesic ODE are $\gamma(0)=\sigma$ and $\dot{\gamma}(0) = \vb{u}$. The point $\gamma(1)\in M^{n}$ is assigned coordinates $(\sigma,\vb{u})$, and in this way one coordinates a tubular neighborhood of $\Sigma$. Note that the manifold of initial data for the geodesics is precisely the normal bundle $(T\Sigma)^{\perp}$. The discussion in this paragraph constructs a map $\psi: \mathbb{R} \times (T\Sigma)^{\perp} \to M^{n}$ given by the geodesic $\psi: (\lambda,\sigma,\vb{u}) \mapsto \gamma(\lambda)$ with initial data $(\sigma,\vb{u})$. The coordinate map is given by the assignment $\gamma(1) \in M^{n} \leftrightarrow (\sigma,\vb{u}) \in (T\Sigma)^{\perp}$.

\begin{figure}
\centering
\includegraphics[width=0.6\textwidth]{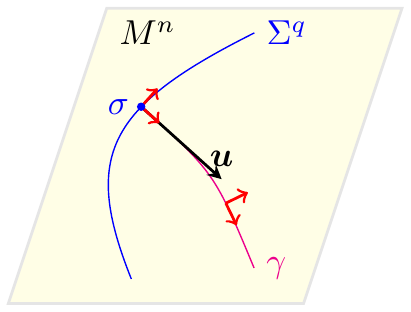}
\caption{At a point $\sigma$ in the embedded submanifold $\Sigma^{q} \hookrightarrow M^{n}$ there is a Darboux frame (in red). This frame is parallel transported along a geodesic $\gamma$ (in magenta) with initial velocity $\vb{u}$ that is orthogonal to $T_{\sigma}\Sigma^{q}$. In this way an orthonormal frame (in red) is constructed along the geodesic. By repeating the construction for every orthogonal direction and for every point in a neighborhood of $\sigma$, one constructs an orthonormal frame field in a tubular neighborhood.}
\label{fig:cartan-method}
\end{figure}

In a neighborhood $U_{\sigma} \subset \Sigma$ of a point $\sigma \in \Sigma$ we construct a Darboux frame, an orthonormal frame where the first $q$ vectors are tangent to $\Sigma$ and the remaining $l=n-q$ vectors are orthogonal to $\Sigma$, see Figure~\ref{fig:cartan-method}. Note that $T_{\sigma}\Sigma \oplus (T_{\sigma}\Sigma)^{\perp} = T_{\sigma} M^{n}$. This Darboux frame is then parallel transported along all the geodesics normal to $\Sigma$ using the Levi-Civita connection on $M^{n}$. In this way an orthonormal coframe is constructed for $M^{n}$ in a tubular neighborhood of $U_{\sigma} \subset \Sigma$ that can be used to obtain an explicit expression for the metric. Cartan's method uses the Darboux frame as initial data for an ODE that is solved to obtain the orthonormal coframe. Any local orthonormal coframe $\theta^{\mu}$ on $M^{n}$ with associated Levi-Civita connection $\omega_{\mu\nu}= -\omega_{\nu\mu}$ will satisfy the Cartan structural equations for a manifold of constant sectional curvature $k$:
	\begin{subequations}\label{eq:cartan-structure}
		\begin{align}
		\dd{\theta^\mu} &= -\omega_{\mu\nu}\wedge\theta^{\nu} \label{eq:cartan-torsion}\\
		\dd{\omega_{\mu\nu}} &= -\omega_{\mu\lambda}\wedge\omega_{\lambda\nu} + k\, \theta^{\mu}\wedge\theta^{\nu} \label{eq:cartan-curvature}
		\end{align}
	\end{subequations}
By using the geodesic map $\psi: \mathbb{R} \times (T\Sigma)^{\perp} \to M^{n}$, we can pull back $(\theta^{\mu}, \omega_{\nu\rho})$ to $\mathbb{R} \times (T\Sigma)^{\perp}$ and in this way we can relate the coframe at a point $\gamma(1) \in M^{n}$ to a coframe at the associated coordinate $(\sigma,\vb{u})\in (T\Sigma)^{\perp}$. A differential form on the product manifold $\mathbb{R}\times (T\Sigma)^{\perp}$ automatically has a bi-degree. If $\lambda$ is the coordinate on $\mathbb{R}$  then a differential form on $\mathbb{R}\times (T\Sigma)^{\perp}$ is a linear combination of forms of degree $0$ and degree $1$ in $\dd\lambda$.

Next we specify how to construct the coframe by exhibiting the ODE it satisfies. The tangent vector to the curve is $\psi_{*}(\partial/\partial\lambda)$. Our initial data is a Darboux frame on $\Sigma$ so we choose an index convention where the indices $a,b,c,\dotsc$ run from $1,2,\dotsc,q$ and are associated with $(T\Sigma)$, and $i,j,k,\dotsc$ run from $q+1,q+2,...,n$ and are associated with $(T\Sigma)^{\perp}$. The Darboux coframe will be denoted by $(\varphi^{a}, \varphi^{i})$. Since the initial data $\vb{u}$ was orthogonal to $\Sigma$, the extension of the Darboux frame by parallel transport will be required to have the same property: $\theta^{a}(\psi_{*}(\partial/\partial\lambda))=0$ and $\theta^{i}(\psi_{*}(\partial/\partial\lambda)) = u^{i}$. The parallel transport condition is $\omega_{\mu\nu}(\psi_{*}(\partial/\partial\lambda))=0$. These conditions state that the differential forms pulled back to $\mathbb{R} \times (T\Sigma)^{\perp}$ satisfy: $(\psi^{*}\theta^{a})(\partial/\partial\lambda)=0$, $(\psi^{*}\theta^{i})(\partial/\partial\lambda)=u^{i}$, and $(\psi^{*}\omega_{\mu\nu})(\partial/\partial\lambda)=0$. This means that
	\begin{subequations}\label{eq:pull-back}
		\begin{align}
		\psi^{*}\theta^a &= \vartheta^a \label{eq:pull-back-a}\\
		\psi^{*}\theta^i &= \vartheta^i + u^i \dd{\lambda} \label{eq:pull-back-b}\\
		\psi^{*}\omega_{\mu\nu} &= \varpi_{\mu\nu} \label{eq:pull-back-c}
		\end{align}
	\end{subequations}
where the $\vartheta^{\mu}$ and $\varpi_{\mu\nu}$ are differential forms that are degree $0$ in $\dd\lambda$ and  are degree~$1$ in the differentials $\dd\sigma^{a}$ and $\dd u^{i}$.

Next consider the exterior derivative of eq.~\eqref{eq:pull-back-a}, use naturalness of the exterior derivative: $\dd \circ \psi^{*} = \psi^{*} \circ \dd$, and  substitute \eqref{eq:cartan-torsion}. In the left-hand side we obtain
\begin{align*}
\psi^{*}\dd{\theta^a} &= -\psi^{*}\omega_{ab}\wedge\psi^{*}\theta^{b} - \psi^{*}\omega_{ai}\wedge\psi^{*}\theta^{i} \\
&= -\varpi_{ab}\wedge\vartheta^{b} - \varpi_{ai}\wedge(\vartheta^{i} + u^{i}\dd{\lambda})
\end{align*}
The exterior derivative of the right-hand side of \eqref{eq:pull-back-a} gives
\begin{equation*}
d\vartheta^{a}  = \dd\lambda \wedge \frac{\partial \vartheta^{a}}{\partial \lambda} + \dotsb\,,
\end{equation*}
where the ellipsis above denotes terms that are of degree~$0$ in $\dd\lambda$. Picking out the terms of degree~$1$ in $\dd\lambda$ from both sides leads to
	\begin{equation}
	\pdv{\vartheta^a}{\lambda} = \varpi_{ai}u^{i} \label{ode:thetaa}
	\end{equation}

Next we take the exterior derivative of both sides of \eqref{eq:pull-back-b} and we substitute \eqref{eq:cartan-torsion} into the left-hand side.
	\begin{align*}
	 -\varpi_{ij}\wedge(\vartheta^{j} + u^{j}\dd{\lambda}) - \varpi_{ia}\wedge\vartheta^{a} &=
	\dd(\vartheta^{i} + u^{i}\dd{\lambda}) \\
	 -\varpi_{ij}\wedge\vartheta^{j} - u^{j}\varpi_{ij}\wedge\dd{\lambda} - \varpi_{ia}\wedge\vartheta^{a} &= \dd{\vartheta^{i}} + \dd{u^i}\wedge\dd{\lambda}\\
	 u^{j} \dd{\lambda}\wedge\varpi_{ij} + \cdots &= \dd{\lambda}\wedge\pdv{\vartheta^{i}}{\lambda} - \dd{\lambda} \wedge \dd{u^i} + \cdots
	\end{align*}
Picking terms of degree~$1$ in $\dd\lambda$, we arrive at the differential equation for the $\vartheta^{i}$:
	\begin{equation}
	\pdv{\vartheta^{i}}{\lambda} = \dd{u^i} + \varpi_{ij}u^{j} \label{ode:thetai}
	\end{equation}

	We now turn our attention towards obtaining differential equations for the connection forms. From the second structural equation \eqref{eq:cartan-curvature} with ``indices along $\Sigma$'' we have: 	\begin{align}
	\psi^{*}\dd{\omega_{ab}} &= \psi^{*} \left( -\omega_{ac}\wedge\omega_{cb} -\omega_{ai} \wedge \omega_{ib}+ k \theta^{a}\wedge\theta^{b} \right) \nonumber\\
	\dd{\varpi_{ab}} &= -\varpi_{ac}\wedge\varpi_{cb} - \varpi_{ai} \wedge \varpi_{ib} + k \vartheta^{a}\wedge\vartheta^{b} \nonumber \\
	\dd{\lambda}\wedge\pdv{\varpi_{ab}}{\lambda} + \cdots &= \cdots
	\end{align}
	We conclude that
	\begin{equation}
	\pdv{\varpi_{ab}}{\lambda} = 0 \label{conn-ab}
	\end{equation}
	
	For the ``mixed indices'' connection forms we have
	\begin{align}
		\psi^{*}\dd{\omega_{ai}} &= \psi^{*} \left( -\omega_{ab}\wedge\omega_{bi}-\omega_{aj}\wedge\omega_{ji} + k\theta^{a}\wedge\theta^{i} \right) \nonumber\\
		\dd{\varpi_{ai}} &= -\varpi_{ab}\wedge\varpi_{bi} -\varpi_{aj}\wedge\varpi_{ji} + k\vartheta^{a}\wedge(\vartheta^{i} + u^{i}\dd{\lambda}) \nonumber\\
		\dd{\lambda}\wedge\pdv{\varpi_{ai}}{\lambda} + \cdots &= -k u^{i} \dd{\lambda}\wedge\vartheta^{a} + \cdots
		\end{align}
	Picking out the degree~$1$ terms in $\dd\lambda$ we find
	\begin{align}
	\pdv{\varpi_{ai}}{\lambda} = -ku^{i}\vartheta^{a} \label{ode:conn-ai}
	\end{align}
	
	For connection forms with ``normal bundle indices'':
	\begin{align}
	\psi^{*}\dd{\omega_{ij}} &= \psi^{*} \left( -\omega_{ik}\wedge\omega_{kj} - \omega_{ia}\wedge \omega_{aj} + k \theta^{i}\wedge\theta^{j} \right)\nonumber\\
	\dd{\varpi_{ij}} &= -\varpi_{ik}\wedge\varpi_{kj} -\varpi_{ia} \wedge \varpi_{aj}+ k (\vartheta^{i}+u^{i}\dd\lambda)\wedge(\vartheta^{j} + u^{j}\dd{\lambda}) \nonumber \\
	&= -\varpi_{ik}\wedge\varpi_{kj} -\varpi_{ia} \wedge \varpi_{aj} + k\vartheta^{i}\wedge\vartheta^{j} + ku^{j}\vartheta^{i}\wedge\dd{\lambda} + ku^{i}\dd{\lambda}\wedge\vartheta^{j} \nonumber\\
	\dd{\lambda}\wedge\pdv{\varpi_{ij}}{\lambda} + \cdots &= \dd{\lambda}\wedge k (u^{i}\vartheta^{j} - u^{j}\vartheta^{i}) + \cdots
	\end{align}
Picking out the terms of degree~$1$ in $\dd\lambda$ we find
	\begin{equation}
	\pdv{\varpi_{ij}}{\lambda} = k\qty(u^{i}\vartheta^{j} - u^{j}\vartheta^{i}) \label{ode:conn-ij}
	\end{equation}
	
Cartan observed that these first order ODEs can be combined into second order ODEs for the coframe.
If we differentiate equation~(\ref{ode:thetaa}) with respect to $\lambda$ and substitute equation~(\ref{ode:conn-ai}) we obtain
	\begin{align}
	\pdv[2]{\vartheta^{a}}{\lambda} &= \pdv{\varpi_{ai}}{\lambda}u^{i} \nonumber\\
	&= -k u^{i}u_{i}\vartheta^{a} \nonumber\\
	&= -k \norm{u}^2 \vartheta^{a} \label{ode:conn-accel-tang}
	\end{align}

Now if we differentiate equation~(\ref{ode:thetai}) with respect to $\lambda$, and use the results of equation~(\ref{ode:conn-ij}) we obtain
	\begin{align}
	\pdv[2]{\vartheta^{i}}{\lambda} &= \pdv{\varpi_{ij}}{\lambda}u^{j} \nonumber\\
	&= k(u^{i}\vartheta^{j} - u^{j}\vartheta^{i})u^{j} \nonumber\\
	&= -k(u^2\delta^{i}_{j} - u^{i}u^{j})\vartheta^{j} \label{ode:accel-soln-thetai}
	\end{align}

Equations \eqref{ode:conn-accel-tang} and \eqref{ode:accel-soln-thetai} are a closed set of second order ODEs that determine the coframe once the initial data are specified. To understand the initial data, it suffices to analyze the problem in flat space with $\Sigma^{q}$ a $q$-plane. In this case the map $\psi$ will be given by $x^{a} = \sigma^{a}$ and $x^{i} = \lambda u^{i}$, which leads to $\dd x^{a}= \dd\sigma^{a}$ and $\dd x^{i} = \lambda\, \dd u^{i} + u^{i}\, \dd\lambda$. We have that $\vartheta^{a}= \dd\sigma^{a}$ and $\vartheta^{i} = \lambda\, \dd u^{i}$. If we go back to the general case we see that $\eval{\vartheta^a}_{\lambda=0} =  \varphi^{a}$ and $\eval{\vartheta^{i}}_{\lambda=0}=0$. Next we need the first derivatives of the coframes. From equation \eqref{ode:thetaa}, we see that
$\eval{\pdv{\vartheta^a}{\lambda}}_{\lambda=0} = \eval{\varpi_{ai} u^i}_{\lambda=0} = K_{abi} u^i \varphi^b$ by the definition of the second fundamental form. Using \eqref{ode:thetai} we see that $\eval{\pdv{\vartheta^{i}}{\lambda}}_{\lambda=0}= du^{i} + \pi_{ij}u^{j}= Du^{i}$ where $\pi_{ij}= \eval{\varpi_{ij}}_{\lambda=0}$ is the orthogonal connection on the normal bundle. We have determined the initial conditions for our second order ODEs for the coframe.

	We now use these to solve equation~(\ref{ode:conn-accel-tang}) for the unknown one-forms $\vartheta^{a}$. We split this into the two cases of positive ($k>0$) and negative ($k<0$) sectional curvature. For the case of $k=0$ the submanifold is flat and we refer the reader to the calculation in~\cite{alvarez:emergentflat}.
	
	In the case of $k>0$, (\ref{ode:conn-accel-tang}) becomes
	\begin{equation}
	\pdv[2]{\vartheta^{a}}{\lambda} + k \norm{u}^2 \vartheta^{a} = 0
	\end{equation}
	
	This equation has the solution
	\begin{align}
	\vartheta^{a} &= \eval{\vartheta^{a}}_{\lambda=0} \cos(\sqrt{k\norm{u}^2}\lambda) + \frac{1}{\sqrt{k\norm{u}^2}} \eval{\pdv{\vartheta^{a}}{\lambda}}_{\lambda=0} \sin(\sqrt{k\norm{u}^2}\lambda) \nonumber\\
	&= \varphi^{a} \cos(\sqrt{k\norm{u}^2}\lambda) + K_{abi}u^{i}\varphi^{b} \frac{\sin(\sqrt{k\norm{u}^2}\lambda)}{\sqrt{k\norm{u}^2}}\nonumber\\
	&= \qty[\cos(\sqrt{k\norm{u}^2}\lambda) \delta^{a}_{b} + K_{abi}u^{i} \frac{\sin(\sqrt{k\norm{u}^2}\lambda)}{\sqrt{k\norm{u}^2}}]\varphi^{b}  \label{ode:accel-soln-kpos}
	\end{align}
	
	Likewise, in the case of $k<0$, the solution is:
	\begin{equation}
	\vartheta^{a} = \qty[\cosh(\sqrt{-k\norm{u}^2}\lambda)\delta^{a}_{b} + K_{abi}u^{i}\frac{\sinh(\sqrt{-k\norm{u}^2}\lambda)}{\sqrt{-k\norm{u}^2}}]\varphi^{b}
	\end{equation}

To solve equation \eqref{ode:accel-soln-thetai} it is useful to define $P_{L}^{ij}= u^{i}u^{j}/\lVert \vb{u} \rVert^{2}$, the orthogonal projector along the velocity $\vb{u}$, and $P_{T}^{ij} =\delta^{ij}- 	u^{i}u^{j}/\lVert \vb{u} \rVert^{2}$, the orthogonal projector transverse to the velocity $\vb{u}$. Note that \eqref{ode:accel-soln-thetai} may be written as
	\begin{equation}
	\pdv[2]{\vartheta^{i}}{\lambda} = -k \norm{u}^2 {P_T}^{i}_{j}\,\vartheta^{j}
	\end{equation}	
We can decompose the $\vartheta^{i}$ frames as
	\begin{subequations}\label{eq:proj}
		\begin{align}
		\vartheta_L &\equiv P_L\vartheta \\
		\vartheta_T &\equiv P_T\vartheta \\
		\vartheta &= \vartheta_L + \vartheta_T
		\end{align}
	\end{subequations}
We then have the uncoupled differential equations
	\begin{subequations}
		\begin{align}
		\pdv[2]{\vartheta_{L}}{\lambda} &= 0 \label{ode:theta-L}\\
		\pdv[2]{\vartheta_{T}}{\lambda} &= -k \norm{u}^2 \vartheta_{T} \label{ode:theta-T}
		\end{align}
	\end{subequations}
Integrating (\ref{ode:theta-L}) and using the initial conditions gives
	\begin{equation}
	\vartheta_L^{i} = \lambda\, P_L (Du)^{i}
	\label{eq:theta-long}
	\end{equation}
Integrating (\ref{ode:theta-T}) gives
	\begin{subequations}
		\begin{numcases}{\vartheta_{T}^{i} =}
		\frac{\sin(\sqrt{k\norm{u}^2}\;\lambda)}{\sqrt{k\norm{u}^2}}\, P_T (Du)^{i} & $k > 0$ \\
		\frac{\sinh(\sqrt{-k\norm{u}^2}\;\lambda)}{\sqrt{-k\norm{u}^2}} P_T (Du)^{i} & $k < 0$
		\label{eq:theta-perp}
		\end{numcases}
	\end{subequations}

In the ensuing discussion we consider the $k<0$ case.
We can simplify the notation a bit by making the observation that  $\sqrt{- k \norm{u}^2} = \abs{k}^{1/2}\norm{u}$. To get our orthonormal coframe we need to take $\lambda=1$ and we obtain
	\begin{align}
	\vartheta^{a} &= \cosh(\abs{k}^{1/2}\norm{u}) \qty[\delta^{a}{}_{b} + \frac{\tanh(\abs{k}^{1/2}\norm{u})}{\abs{k}^{1/2}\norm{u}}\, u^{j} K_{abj}]\varphi^{b} \nonumber\\
	&= \cosh(\abs{k}^{1/2}\norm{u})\qty[\delta^{a}{}_{b} + \frac{\tanh(\abs{k}^{1/2}\norm{u})}{\abs{k}^{1/2}}\, \hat{u}^{j} K_{abj}]\varphi^{b} 
	\label{eq:theta-a}
	\end{align}
The corresponding part of the volume element is:
	\begin{equation}
	\vartheta^{1} \wedge \cdots \wedge \vartheta^{q} = \qty[\cosh(\abs{k}^{1/2}u)]^{q}\det(I + \frac{\tanh(\abs{k}^{1/2}u)}{\abs{k}^{1/2}} \;\vb{\hat{u}}\vdot \vb{K})\vb*{\zeta}_{\Sigma}
	\end{equation}

Next we observe that $P_T \vartheta \perp P_L \vartheta$. 	The ``normal piece'' of the volume element (corresponding to the $\vartheta^i$) is:
	\begin{equation}
	\vartheta^{q+1} \wedge \cdots \wedge \vartheta^{n} = \qty(\frac{\sinh(\abs{k}^{1/2}u)}{\abs{k}^{1/2} u})^{l-1} Du^{q+1}\wedge Du^{q+2}\cdots \wedge Du^{q+l}
	\end{equation}	
where $l = n - q = \text{codim}\ \Sigma$. Combining these two gives the full volume element (for a space with $k<0$) as:
	\begin{align}
	&\vartheta^{1}\wedge\cdots\wedge\vartheta^{q}\wedge\vartheta^{q+1}\wedge\cdots\wedge\vartheta^{n} = \qty[\cosh(\abs{k}^{1/2} u)]^{q} \qty(\frac{\sinh(\abs{k}^{1/2}u)}{\abs{k}^{1/2}u})^{l-1} \nonumber\\
	&\quad\times\det(I + \frac{\tanh(\abs{k}^{1/2}u)}{\abs{k}^{1/2}} \vb{\hat{u}}\vdot\vb{K}) \vb*{\zeta}_{\Sigma}\wedge Du^{q+1} \wedge \cdots \wedge Du^{n} \nonumber
	\end{align}
Remembering that $\vb*{\zeta}$ is already is already of maximal degree $q$ in $d\sigma^{a}$, and that the normal bundle connection $\pi_{ij}$ is a $1$-form on $\Sigma$ (only $\dd\sigma^{a}$ appear), we see that the volume element may be rewritten as
	\begin{align}
	&\vartheta^{1}\wedge\cdots\wedge\vartheta^{q}\wedge\vartheta^{q+1}\wedge\cdots\wedge\vartheta^{n} = \qty[\cosh(\abs{k}^{1/2} u)]^{q} \qty(\frac{\sinh(\abs{k}^{1/2}u)}{\abs{k}^{1/2}u})^{l-1} \nonumber\\
	&\quad\times\det(I + \frac{\tanh(\abs{k}^{1/2}u)}{\abs{k}^{1/2}} \; \vb{\hat{u}}\vdot\vb{K}) \vb*{\zeta}_{\Sigma}\wedge \dd u^{q+1} \wedge \cdots \wedge \dd u^{n}
	\label{eq:k-vol-element}
	\end{align}
in analogy to the flat-space case discussed in the companion paper.

	Similarly, for $k>0$:
	\begin{align}
	&\vartheta^{1}\wedge\cdots\wedge\vartheta^{q}\wedge\vartheta^{q+1}\wedge\cdots\wedge\vartheta^{n} = \qty[\cos(\abs{k}^{1/2} u)]^{q} \qty(\frac{\sin(\abs{k}^{1/2}u)}{\abs{k}^{1/2}u})^{l-1} \nonumber\\
	&\quad\times\det(I + \frac{\tan(\abs{k}^{1/2}u)}{\abs{k}^{1/2}} \;\vb{\hat{u}}\vdot\vb{K}) \vb*{\zeta}_{\Sigma}\wedge \dd{u^{q+1}} \wedge \cdots \wedge \dd{u^{n}}
	\end{align}
	
	\section{Multipole expansion for the action (energy) of a tube}
	
The action (energy) of a tube is given by
\begin{equation}
    E = \int_{\Sigma} \vb*{\zeta}_{\Sigma} (\sigma) \left( 
    \int_{(T_{\sigma}\Sigma)^{\perp}} u(\sigma, \vb*{\nu})\; \det(I + \vb*{\nu}\vdot\vb{K})\;
    d^{l}\nu \right),
    \label{eq:total-energy}
\end{equation}
where $u(\sigma, \vb*{\nu})$ is the energy density.	Here $\sigma$ is a point on $\Sigma^{q}$ and $\nu^{i}$ are Cartesian coordinates on the normal bundle fiber $(T_{\sigma}\Sigma)^{\perp}$. The energy density has a multipole expansion given by
\begin{equation*}
u(\sigma, \vb*{\nu}) = \sum_{j=0}^{\infty} \sum_{k_{1},\dotsc,k_{j}}
u^{(j)}_{k_{1}\cdots k_{j}}(\sigma,\norm{\vb*{\nu}})\;
\mathcal{Y}^{j}_{k_{1}\cdots k_{j}}(\hat{\vb*{\nu}}),
\end{equation*}
where $\mathcal{Y}^{j}_{k_{1}\cdots k_{j}}$ is a Cartesian spherical harmonic defined in \cite{alvarez:emergentflat},
$ u^{(j)}_{k_{1}\cdots k_{j}}(\sigma,\norm{\vb*{\nu}})$ is a Cartesian multipole expansion coefficient\footnote{This Cartesian multipole expansion moment is a symmetric traceless tensor in the $k_{1}\cdots k_{j}$ indices.}, $\sigma$ is a point on the submanifold $\Sigma$, $\vb*{\nu}\in(T_{\sigma}\Sigma)^{\perp}$ is a vector in the normal bundle, and $\vb*{\zeta}_{\Sigma}$ is the volume element on the submanifold. 
	It was shown~\cite{alvarez:emergentflat} that the contribution to the energy of a tube in $\mathbb{E}^n$ is $E = \sum_{j=0}^{\infty} E^{(j)}$ where the contribution from the $2^{j}$-pole is:
	\begin{equation}
	E^{(j)} = \int_{\Sigma} \vb*{\zeta}_{\Sigma} \int_{(T_{\sigma}\Sigma)^{\perp}} u^{(j)}_{k_{1}\cdots k_{j}}(\sigma,\norm{\vb*{\nu}})\, \hat{\nu}^{k_{1}}\hat{\nu}^{k_{2}}\cdots\hat{\nu}^{k_{j}} \det(I + \vb*{\nu}\vdot\vb{K})\dd[l]{\nu}
	\end{equation}
A surprising result of the companion paper is that only $E^{(j)}$ with $j \le q$ contribute to the energy.
	
	The derivation of the multipole expansion in \cite{alvarez:emergentflat} generalizes to the constant curvature manifold $M^{n}$ if we replace the Euclidean volume element by the constant curvature volume element~\eqref{eq:k-vol-element} and obtain\footnote{Here we only present the results for negative curvature. For the positive curvature results, just replace the hyperbolic functions by the corresponding circular functions.}:
	\begin{align}
	E^{(j)} &= \int_{\Sigma} \vb*{\zeta}_{\Sigma}\int_{(T_{\sigma}\Sigma)^{\perp}} u^{(j)}_{k_{1}\cdots k_{j}}(\sigma,\norm{\vb*{\nu}})\, \hat{\nu}^{k_{1}}\hat{\nu}^{k_{2}}\cdots\hat{\nu}^{k_{j}} \nonumber\\
	&\times\det(I + \frac{\tanh(\abs{k}^{1/2}\nu)}{\abs{k}^{1/2}\nu}\vb*{\nu}\vdot\vb{K})\qty(\cosh(\abs{k}^{1/2}\nu))^{q} \qty(\frac{\sinh(\abs{k}^{1/2}\nu)}{\abs{k}^{1/2}\nu})^{l-1}\; \dd[l]{\nu}
	\end{align}
	
	In order to perform this integral, we need to expand the determinant using the identity
	\begin{equation}
	\det(I+tS) = \sum_{m=0}^{n} \frac{t^m}{m!} \,\delta_{i_1 \cdots i_m}^{j_1 \cdots j_m}\, \tensor{S}{^{i_1}_{j_1}} \tensor{S}{^{i_2}_{j_2}} \cdots \tensor{S}{^{i_m}_{j_m}}\,,
	\end{equation}
where $S$ is a symmetric matrix~\cite{alvarez:emergentflat}.
	Doing so turns the integral over the normal bundle into an integral over the \mbox{$(l-1)$-sphere} and the radial direction, with the $\nu = \norm{\vb*{\nu}}$ acting as the radial coordinate:
	\begin{align}
	E^{(j)} &= \int_{\Sigma}\vb*{\zeta}_{\Sigma} \sum_{r=0}^{q} \int_{\nu=0}^{\infty} \int_{S^{l-1}} u^{(j)}_{k_1 \cdots k_j}(\sigma,\nu) \frac{\nu^r}{r!}\delta^{b_1 \cdots b_r}_{a_1 \cdots a_r} \tensor{K}{^{a_1}_{b_1 i_1}}(\sigma) \tensor{K}{^{a_2}_{b_2 i_2}}(\sigma) \cdots \tensor{K}{^{a_r}_{b_r i_r}}(\sigma) \nonumber\\
	&\quad\times\qty(\cosh \abs{k}^{1/2}\nu)^q \qty(\frac{\sinh \abs{k}^{1/2}\nu}{\abs{k}^{1/2}\nu})^{l-1} \qty(\frac{\tanh \abs{k}^{1/2}\nu}{\abs{k}^{1/2}\nu})^r \nonumber \\
	&\quad\times \hat{\nu}^{k_1}\hat{\nu}^{k_2}\cdots\hat{\nu}^{k_j}\hat{\nu}^{i_1}\hat{\nu}^{i_2}\cdots\hat{\nu}^{i_r}\nu^{l-1}\dd{\nu} \dd{\mathrm{vol}_{S^{l-1}}}\nonumber \\
	&=\int_{\Sigma}\vb*{\zeta}_{\Sigma} \sum_{r=0}^{q} \int_{\nu=0}^{\infty} \int_{S^{l-1}} u^{(j)}_{k_1 \cdots k_j}(\sigma,\nu) \frac{\nu^r}{r!}\delta^{b_1 \cdots b_r}_{a_1 \cdots a_r} \tensor{K}{^{a_1}_{b_1 i_1}}(\sigma) \tensor{K}{^{a_2}_{b_2 i_2}}(\sigma) \cdots \tensor{K}{^{a_r}_{b_r i_r}}(\sigma) \nonumber\\
	&\quad\times \qty(\cosh \abs{k}^{1/2}\nu)^{q-r} \qty(\frac{\sinh \abs{k}^{1/2}\nu}{\abs{k}^{1/2}\nu})^{r+l-1} \nonumber \\
	&\quad\times \hat{\nu}^{k_1}\hat{\nu}^{k_2}\cdots\hat{\nu}^{k_j}\hat{\nu}^{i_1}\hat{\nu}^{i_2}\cdots\hat{\nu}^{i_r}\nu^{l-1}\dd{\nu} \dd{\mathrm{vol}_{S^{l-1}}}\nonumber \\
	&= \int_{\Sigma} \vb*{\zeta}_{\Sigma} \sum_{r \in \mathcal{R}} \int_{0}^{\infty} \dd{\nu} u^{(j)}_{k_1\cdots k_j}(\sigma,\nu) \frac{\nu^{r+l-1}}{r!} \delta^{b_1\cdots b_r}_{a_1\cdots a_r} \tensor{K}{^{a_1}_{b_1 i_1}}(\sigma) \tensor{K}{^{a_2}_{b_2 i_2}} \cdots \tensor{K}{^{a_r}_{b_r i_r}}(\sigma) \nonumber \\
	&\quad\times \qty(\cosh \abs{k}^{1/2}\nu)^{q-r} \qty(\frac{\sinh \abs{k}^{1/2}\nu}{\abs{k}^{1/2}\nu})^{r+l-1} V_{l-1}(S^{l-1})C_{j+r}\mathcal{W}^{k_1\cdots k_j i_1 \cdots i_r} \label{multienergy-k}
	\end{align}
	We follow the same treatment here as in \cite{alvarez:emergentflat}. The $C_{j+r}$ are normalization constants, while $\mathcal{W}^{k_1 \cdots k_j i_1 \cdots i_r}$ is an expression with $(j+r-1)!!$ terms constructed from all possible Wick contractions on pairs of Kronecker deltas $\delta^{k_1 k_2}\cdots\delta^{k_j i_1}\cdots\delta^{i_{r-1} i_r}$. The product $C_{j+r}\mathcal{W}^{k_1\cdots k_j i_1 \cdots i_r}$ arises as a result of averaging over $S^{l-1}$. Note that $u^{(j)}_{k_1\cdots k_j}$ is symmetric and traceless, so any term containing $\delta^{kk'}$ will vanish. In order to have a nonzero contribution, a ``$k$'' index must contract with an ``$i$'' index. Thus, the number of terms that are nonzero must have $r-j$ even and we conclude $r-j=2s$ where $0 \leq s \leq \floor{(q-j)/2}$. Putting all this together, equation \eqref{multienergy-k} reads
	\begin{align}
	E^{(j)} &= \int_{\Sigma} \vb*{\zeta}_{\Sigma} \sum_{s=0}^{\floor{(q-j)/2}}\int_{0}^{\infty} \dd{\nu} u^{(j)}_{k_1\cdots k_j}(\sigma,\nu) \frac{\nu^{2s+j+l-1}}{(j+2s)!}\delta^{b_1 \cdots b_{j+2s}}_{a_1 \cdots a_{j+2s}} \nonumber \\
	&\quad\times \tensor{K}{^{a_1}_{b_1 i_1}}(\sigma) \tensor{K}{^{a_2}_{b_2 i_2}}(\sigma) \cdots\tensor{K}{^{a_{j+2s}}_{b_{j+2s}}_{i_{j+2s}}} V_{l-1}(S^{l-1}) C_{2(j+s)}\mathcal{W}^{k_1\cdots k_j i_1 \cdots i_{j+2s}} \nonumber \\
	&\quad\times \qty(\cosh \abs{k}^{1/2}\nu)^{q-j-2s} \qty(\frac{\sinh \abs{k}^{1/2}\nu}{\abs{k}^{1/2}\nu})^{j+2s+l-1} \label{multipole-sphereavg}
	\end{align}
	
	It is straightforward to extract the radial moments from this expression. Define the radial $2^j$-pole moment as
	\begin{align}
	\mu_{k_1 \cdots k_j, 2s+j}^{(j)} (\sigma) &= V_{l-1}(S^{l-1}) \int_{0}^{\infty} \dd{\nu} \nu^{2s+j+l-1} u_{k_1 \cdots k_j}^{(j)}(\sigma,\nu) \nonumber\\
	&\quad\times \qty(\cosh \abs{k}^{1/2}\nu)^{q-j-2s} \qty(\frac{\sinh \abs{k}^{1/2}\nu}{\abs{k}^{1/2}\nu})^{2s+j+l-1}\nonumber \\
	&= V_{l-1}(S^{l-1}) \int_{0}^{\infty} \dd{\nu} u_{k_1 \cdots k_j}^{(j)}(\sigma,\nu) \nonumber\\
	&\quad\times  \qty(\cosh \abs{k}^{1/2}\nu)^{q-j-2s} \qty(\frac{\sinh \abs{k}^{1/2}\nu}{\abs{k}^{1/2}})^{2s+j+l-1} \label{multipole-moment}
	\end{align}
	
	This can be put into a more convenient form if we define the radius of curvature $\rho \equiv 1/\sqrt{\abs{k}}$. We can then rewrite the expression above as:
	\begin{align}
	\mu^{(j)}_{k_1 \cdots k_j, 2s + j}(\sigma) &= V_{l-1}\qty(S^{l-1}) \int_{0}^{\infty} \dd{\nu} u^{(j)}_{k_1 \cdots k_j}(\sigma,\nu) \nonumber \\
	&\quad\times \qty(\cosh \frac{\nu}{\rho})^{q-j-2s} \qty(\rho \sinh \frac{\nu}{\rho})^{2s+j+l-1} 
	\label{multipole-moment-rho}
	\end{align}
	
	The total number of nonzero terms in $\mathcal{W}^{k_1\cdots k_j i_1 \cdots i_{j+2s}}$ is $\frac{(j+2s)!}{(2s)!} \times (2s-1)!! = \frac{(j+2s)!}{2^s s!}$. The contribution to the multipole expansion we obtained in equation \ref{multipole-sphereavg} is then:
	\begin{align}
	E^{(j)} &= \int_{\Sigma} \vb*{\zeta}_{\Sigma} \sum_{s=0}^{\floor{(q-j)/2}} \frac{C_{2(j+s)}}{2^s s!} \mu^{(j)}_{k_1 \cdots k_j, 2s+j}(\sigma)\delta_{a_1 \cdots a_{2s+j}}^{b_1 \cdots b_{2s+j}}\nonumber \\
	&\quad\times \tensor{K}{^{a_1}_{b_1 k_1}}(\sigma)\tensor{K}{^{a_2}_{b_2 k_2}}(\sigma)\cdots\tensor{K}{^{a_{2s+j}}_{b_{2s+j} k_{2s+j}}}(\sigma) 
	\end{align}
	Using the Gauss equation, $\tensor{R}{_{abcd}} = \tensor{K}{_{aci}} \tensor{K}{_{bdi}} - \tensor{K}{_{adi}}\tensor{K}{_{bci}}$, we can write this in terms of intrinsic curvature terms:
	\begin{align}
	E^{(j)} &= \int_{\Sigma} \vb*{\zeta}_{\Sigma} \sum_{s=0}^{\floor{(q-j)/2}} \frac{C_{2(j+s)}}{2^s s!} \mu^{(j)}_{k_1 \cdots k_j, 2s+j}(\sigma) \delta^{b_1 \cdots b_{2s+j}}_{a_1 \cdots a_{2s+j}} \nonumber \\
	&\quad\times\tensor{K}{^{a_1}_{b_1}_{k_1}}(\sigma)\tensor{K}{^{a_2}_{b_2}_{k_2}}(\sigma)\cdots\tensor{K}{^{a_j}_{b_j}_{k_j}}(\sigma) \nonumber\\
	&\quad\times \frac{1}{2^s}\tensor{R}{^{a_{j+1}}^{a_{j+2}}_{b_{j+1}}_{b_{j+2}}}\cdots \tensor{R}{^{a_{j+2s-1}}^{a_{j+2s}}_{b_{j+2s-1}}_{b_{j+2s}}}
	\end{align}
	Finally, introducing the extrinsic curvature 1-forms $\kappa_{ai} = K_{abi}\theta^b$, and intrinsic curvature 2-forms $\Omega_{ab} = \frac{1}{2} R_{abcd}\theta^c \wedge \theta^d$ we can write the expression in terms of local geometric data associated with $\Sigma^{q}$:
	\begin{align}
	E^{(j)} &= \sum_{s=0}^{\floor{(q-j)/2}} \frac{C_{2(j+s)}}{2^s s!}  \int_\Sigma \mu^{(j)}_{k_1\cdots k_j,2s+j}(\sigma) \nonumber \\
	&\quad\times \tensor{\kappa}{_{b_1}^{k_1}}\wedge\cdots\wedge\tensor{\kappa}{_{b_j}^{k_j}}\wedge\Omega_{a_1 a_2}\wedge\cdots\wedge\Omega_{a_{2s-1} a_{2s}}\wedge\vb*{\zeta}_{\Sigma}^{b_1 \cdots b_j a_1 \cdots a_{2s}}
	\end{align}
	This expression is formally identical to the flat-space expression in \cite{alvarez:emergentflat} except that the multipole moments are given by \eqref{multipole-moment-rho}.

Of importance for us is the formula for the energy in the spherically symmetric case:
Collating everything, we  have the monopole contribution to the energy
\begin{equation}
    E^{(0)} = \sum_{s=0}^{\lfloor q/2 \rfloor} C_{2s}\; \int_{\Sigma} 
    \mu^{(0)}_{2s}(\sigma)\; \mathcal{K}_{2s}(\Sigma)\; \vb*{\zeta}_{\Sigma}\,,
    \label{eq:gen-Weyl}
\end{equation}
	where
	\begin{equation}
	\mu^{(0)}_{2s}(\sigma) = V_{l-1}\qty(S^{l-1}) \int_{0}^{\infty} \dd{\nu}	 \qty(\cosh \frac{\nu}{\rho})^{q-2s} \qty(\rho \sinh \frac{\nu}{\rho})^{2s+l-1} 
	\; u^{(0)}(\sigma,\nu)\,,
	\label{eq:monopole-moment}
	\end{equation}
and the Lovelock Lagrangians (Lipschitz-Killing curvatures) are defined by
\begin{equation}
    \begin{split}
    \curv_{2r}(\Sigma)\; \dual_{\Sigma} &= \frac{1}{4^{r}\, r!}\; 
    \delta_{a_{1}\dotsm a_{2r}}^{b_{1}\dotsm
    b_{2r}}\; R_{a_{1}a_{2}b_{1}b_{2}} \dotsm
    R_{a_{2r-1}a_{2r}b_{2r-1}b_{2r}}\, \dual_{\Sigma}\,, \\
    &= \frac{1}{2^{r}\, r!}\; \dual^{a_{1}a_{2}\dotsm a_{2r-1}a_{2r}}
    \wedge \Omega_{a_{1}a_{2}} \wedge\dotsb \wedge
    \Omega_{a_{2r-1}a_{2r}}\,.
    \end{split}
    \label{eq:def-curv}
\end{equation}
The expression for $E^{(0)}$ is formally the same as in the flat-space embedding case except for a difference in the definition of the multipole moments~\eqref{eq:monopole-moment} due to the curvature of the embedding space. We introduce $J_{2s}$ to be the Jacobian factor in \eqref{eq:monopole-moment}
\begin{equation}
J_{2s}(\nu) =  V_{l-1}\qty(S^{l-1}) \qty(\cosh \frac{\nu}{\rho})^{q-2s} \qty(\rho \sinh \frac{\nu}{\rho})^{2s+l-1}\,.
\label{eq:def-J}
\end{equation}
In models without $\sigma$ dependence in the moments, such as in the spherically symmetric case, the $p$-brane tension is given by $T_{q} = \mu^{(0)}_{0}$, and the $q$-dimensional Newtonian constant $G_{q}$ and the Planck mass $\MPl_{q}$ are related  by $G^{-1}_{q}=(\MPl_{q})^{q-2} = C_{2}\mu^{(0)}_{2}$. The  cosmological constant, as defined in standard general relativity, is $\Lambda_{q}=T_{q} G_{q}=T_{q}/(\MPl_{q})^{q-2}$ with dimension $L^{-2}$.

	\section{Energy (action) of a spherically symmetric tube}
	\label{sec:energy-action}
	
	We now discuss the effective Lagrangian for a spherically symmetric action (energy) tube embedded in a constant curvature space $\AdS_{n}$.  Namely, the core of the defect is a $q$-dimensional submanifold $\Sigma^{q}$ and its action (energy) density is spherically symmetric about the defect. For example, the  topological defect may be the Nielsen-Olesen vortex embedded into $\text{AdS}_4$.
We assume that near the core the energy density is a smooth function, and that far from the core of the defect the energy density decays exponentially. There are  two distinct categories of length scales in this problem: a generic correlation length $1/m$ associated with the masses $m$  of the excitations, and the length $\rho$ corresponding to the radius of curvature $\rho = \lvert k \rvert^{-1/2}$ of  $\AdS_{n}$. The case  $\rho \gg 1/m$ is essentially the flat-space case considered in the previous paper~\cite{alvarez:emergentflat} because the fields decay before one can see the effects of the curvature. Here we are interested in the case where the radius of curvature is much smaller than the mass correlation length $\rho \ll 1/m$. In examples that will be discussed in a subsequent paper~\cite{Alv:2016c}, we find that the asymptotic behavior of the energy density is given by
	\begin{equation}
	u^{(0)}(\sigma,\norm{\vb*{\nu}}) \sim \frac{C}{\nu^\alpha} \; e^{-\nu/\xi} \quad\text{as } \abs{\nu}\to\infty
	\label{eq:u-approx}
	\end{equation}
where $\nu$ is the radial distance from the core of the defect, $\xi$ is a correlation length that is approximately a multiple of the radius of curvature $\rho$ in the case $\rho \ll 1/m$, $\alpha$ is some usually non-negative exponent\footnote{In several examples of defects embedded in $\AdS_{n}$, we found that for our choice of parameters $\alpha=0$ due to the presence of $\rho$ in the differential equations for the fields. For defects embedded in $\mathbb{E}^{n}$ one finds that $\alpha\ge 0$ in general.}, and $C$ is a constant with dimension $L^{-n+\alpha}= M^{n-\alpha}$. The defect is assumed to have finite ``transverse energy''.

The mechanism we explore assumes that the parameters of the theory are such that $\rho  \ll 1/m$. In this case there is a competition where the exponentially decreasing energy density is challenged by the exponentially increasing volume of the negative constant curvature space. 
The asymptotic growth  of the Jacobian factor \eqref{eq:def-J}  is easily obtained. We observe that as $x \to +\infty$ the hyperbolic functions $\cosh x$ and $\sinh x$ both grow like $e^{x}/2$. This immediately gives us the asymptotic growth of the volume element factor, see \eqref{eq:monopole-moment}, 
\begin{equation}
J_{2s}(\nu) \xrightarrow{\nu\to +\infty\,} V_{l-1}(S^{l-1})
\qty(\frac{1}{2})^{n - 1}\; \rho ^{2s+l-1}\;
e^{(n-1)\nu/\rho}\;,
\end{equation}
where $n=q+l$. The exponential increase in the volume element goes like $e^{(n-1)\nu/\rho}$, while the energy density decreases as $e^{-\nu/\xi}$. This means that the asymptotic behavior of $J_{2s}(\nu)\, u^{(0)}(\sigma,\nu)$ behaves as $e^{-\nu/\xieff}$ where
\begin{equation}
\xieff = \xi\; \frac{1}{1-(n-1)\xi/\rho}\,.
\label{eq:def-xi}
\end{equation}
Convergence of the integral \eqref{eq:monopole-moment} as $\nu\to +\infty$ requires that $\xi < \rho/(n-1)$, and imposing slow exponential decay gives $\xi \lessapprox \rho/(n-1)$. If $\xi > \rho/(n-1)$ then the transverse energy integral diverges. We remind the reader that in the parameter range discussed here, $\xi$ is a roughly proportional to $\rho$.
There are two cases to consider: the first case is where $\xieff \sim \rho$, and the second case is where $\xieff \gg \rho$.

\begin{figure}[t]
\centering
\includegraphics[width=0.75\textwidth]{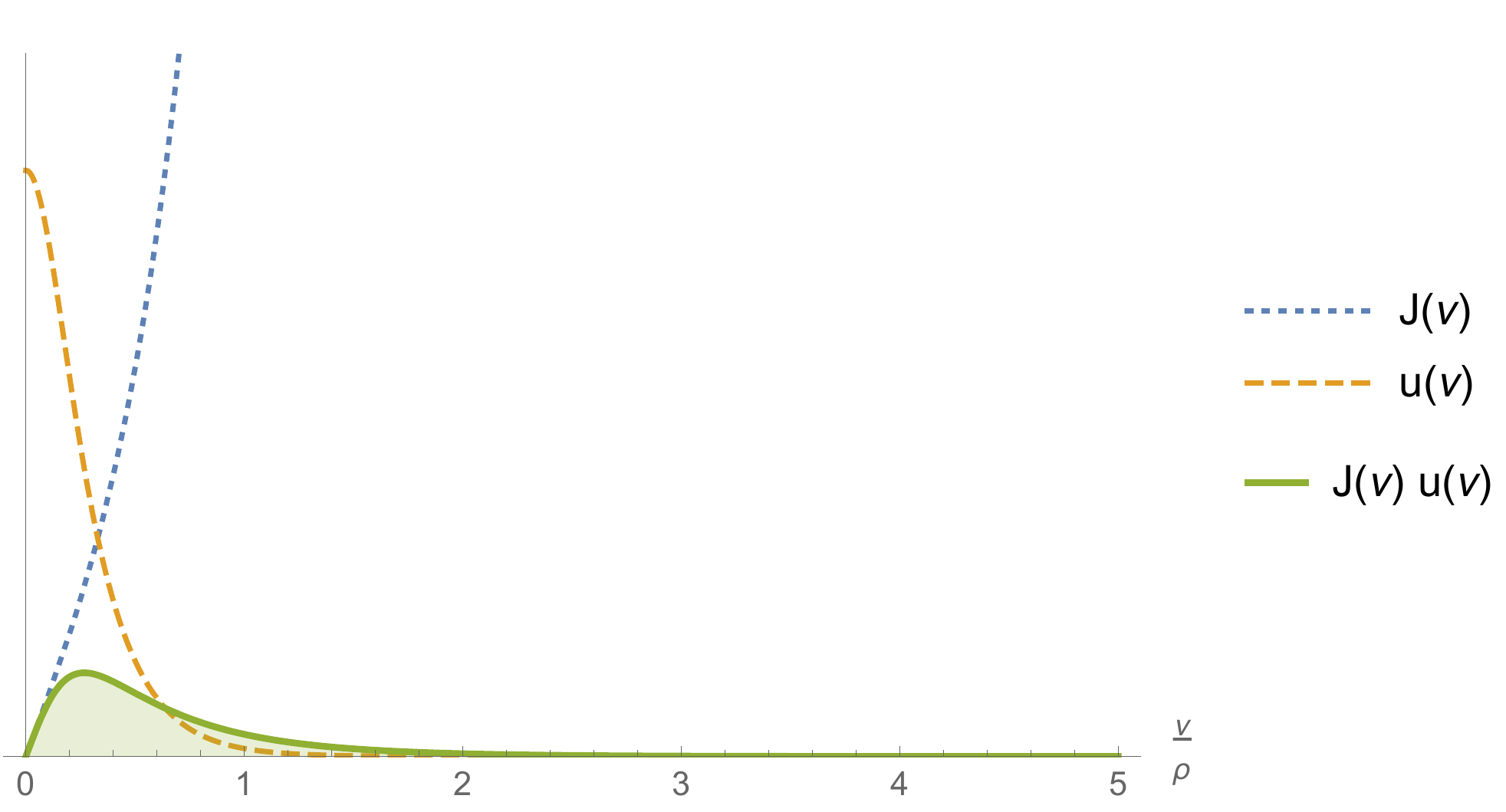}
\caption[xxx1]{Schematic graph (drawn with the same scale as 
Figure~\ref{fig:long-range}) for the computation of the zeroth moment (transverse energy) in the case $n=q+l=2+2$ and short range $\xieff \sim \rho$. The transverse energy is the shaded area under the green curve. Notice that most of the energy comes from a region with $\nu \lessapprox \rho$.}
\label{fig:short-range}
\end{figure}

The case $\xieff \sim \rho$ is illustrated in Figure~\ref{fig:short-range}. Here most of the contribution to the integral \eqref{eq:monopole-moment} comes from the region $\nu < \rho$. The behavior of the moments here are similar to what happens in the Kaluza-Klein case where one picks up a factor of the volume of the compact fiber. The energy will have a factor of $\xieff^{l} \sim \rho^{l}$ which is basically the $l$-volume of a fiber $(T_{\sigma}\Sigma)^{\perp}$ with radius roughly $\rho$.
In these models,  the exponentially decaying tail of the energy density dominates the exponentially increasing volume element. The dominant contribution to the moments is from the core region close to the defect and the results are similar to ones from Kaluza-Klein theory where the compact manifold has a volume $O(\rho^{l})$.

To make a more precise argument, we use a variant of the mean value theorem of integral calculus. From the form of the integrand shown in Figure~\ref{fig:short-range} we note that there is a $\nu = \lambda \rho$ with $\lambda = O(1)$ such that the contribution to the moment integral for $\nu > \lambda \rho$ is negligible. Thus we approximate \eqref{eq:monopole-moment} by
\begin{equation}
\mu^{(0)}_{2s}(\sigma) \approx \int_{0}^{\lambda\rho} \dd\nu\; J_{0}(\nu)\, \left(\rho \tanh\frac{\nu}{\rho}\right)^{2s}\, u^{(0)}(\sigma,\nu)\,.
\label{eq:moment-approx}
\end{equation}
The value of the integral should not be very sensitive to the precise choice of $\lambda$.
The transverse energy contained in the $l$-ball of the radius $\lambda\rho$ in $(T_{\sigma}\Sigma)^{\perp}$ for the case where $\Sigma^{q}$ is a totally geodesic $q$-submanifold~\cite{Alv:2016c} of $\AdS_{n}$ is given by
\begin{align}
\mu^{(0)}_{0}(\sigma) &=\int_{0}^{\lambda\rho} \dd\nu\; J_{0}(\nu)\, u^{(0)}(\sigma,\nu) 
\nonumber \\
&= \rho^{l} \;V_{l-1}(S^{l-1})\int_{0}^{\lambda} \dd z\; (\cosh z)^{q}(\sinh z)^{l-1} \, u^{(0)}(\sigma,\rho z)
= \rho^{l} \;\Eperp(\lambda,\sigma) \,.
\end{align}
Here $\Eperp(\lambda,\sigma)$ is the transverse energy of the radius $\lambda$ ball in the normal fiber of a totally geodesic $q$-submanifold in an $\AdS_{n}$ with sectional curvature $k=-1$, \emph{i.e.}, $\rho=1$. $\Eperp(\lambda)$ should not be very sensitive to the precise choice of $\lambda$. For us the important observation is that the transverse energy is essentially proportional to $\rho^{l}$, which is like the volume of the compactification factor in a Kaluza-Klein scenario.

Applying a mean value-like theorem to \eqref{eq:moment-approx} we find
\begin{equation}
\mu^{(0)}_{2s}(\sigma) =  \rho^{l+2s}\;\Eperp(\lambda,\sigma)\, (\tanh z_{2s})^{2s}\,,
\end{equation}
where $0\le z_{2s} \le \lambda$. In particular, note the ratio
\begin{equation}
\frac{\mu^{(0)}_{2s+2}(\sigma)}{\mu^{(0)}_{2s}(\sigma)} = \rho^{2}\; \left[ \frac{(\tanh z_{2s+2})^{2s+2}}{(\tanh z_{2s})^{2s}} \right].
\label{eq:ratio-short}
\end{equation}
Note that the factor in the square brackets is expected to be $O(1)$.

\begin{figure}[t]
\centering
\includegraphics[width=0.75\textwidth]{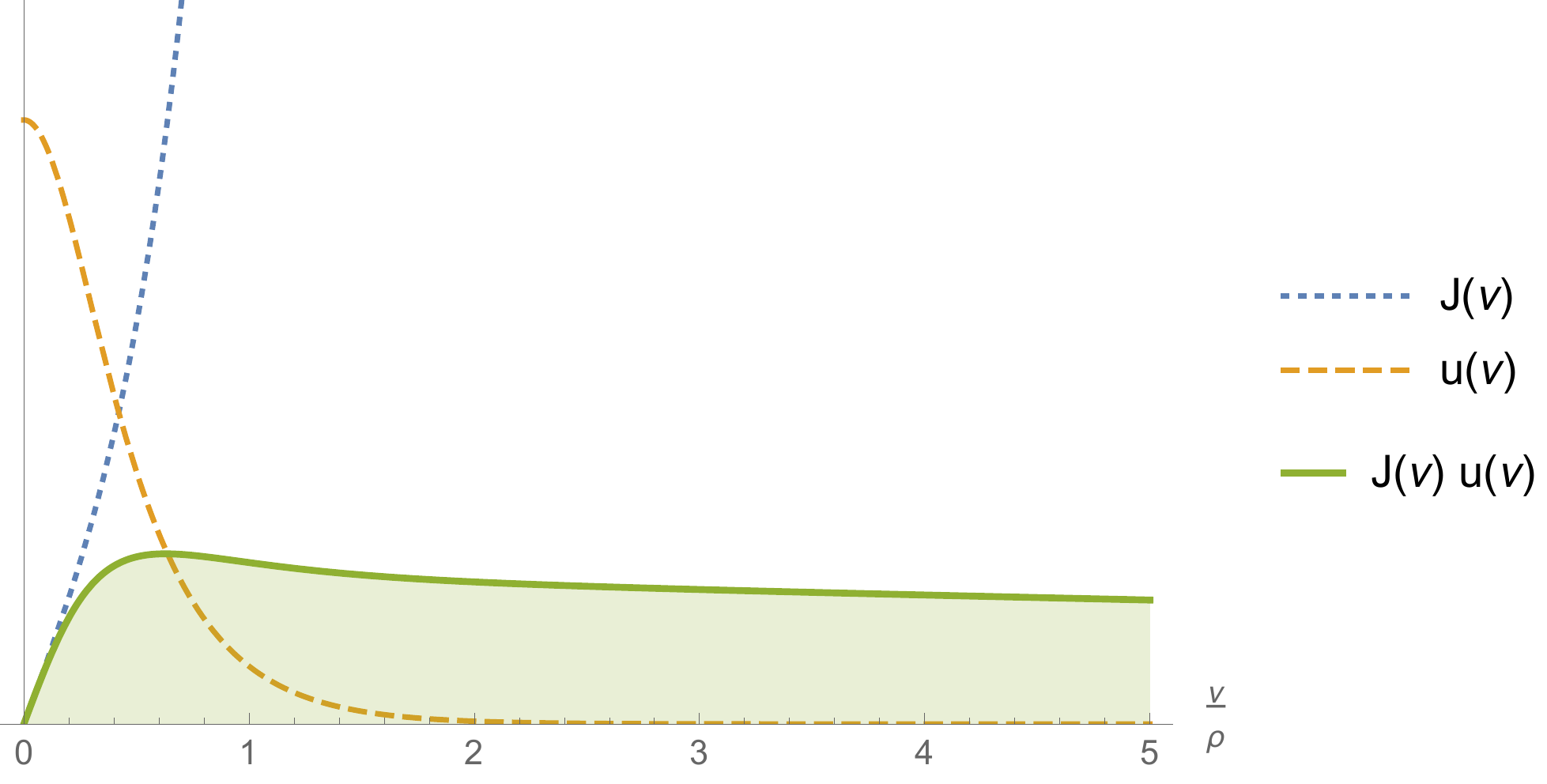}
\caption[yyy1]{Schematic graph (drawn with the same scale as 
Figure~\ref{fig:short-range}) for the computation of the zeroth moment (transverse energy) in the case $n=q+l=2+2$ and longer range $\xieff \gg \rho$. The transverse energy is the shaded area under the green curve. Notice that most of the energy comes from a region with $\nu > \rho$.}
\label{fig:long-range}
\end{figure}

The second scenario, $\xieff \gg \rho$, occurs in models where the asymptotic decay of the energy density slightly overcomes the exponentially increasing volume element, see Figure~\ref{fig:long-range}. In this case the combination $J(\nu) u^{(0)}(\nu)$ has a very long tail and decays very slowly, reaching distances of order $1/m$. The moment integrals can get a  contribution from the slowly decaying tail that is much larger the the core contribution. This is a very different scenario than the Kaluza-Klein one because most of the area under the curve is in the tail $\nu>\rho$. Let $\lambda\rho$ be roughly the value of $\nu$ where the slow exponential decay of $J_{2s}(\nu) u^{(0)}(\nu)$ begins, see Figure~\ref{fig:long-range}. We expect the parameter $\lambda$ to be $O(1)$. In this domain we have that 
\begin{equation}
J_{2s}(\nu)\, u^{(0)}(\nu) \approx J_{2s}(\lambda\rho) \, u^{(0)}(\lambda\rho)\, e^{\lambda\rho/\xieff} \, e^{-\nu/\xieff}\,. 
\end{equation}
In this case we approximate the moment integral by
\begin{align}
\mu^{(0)}_{2s}(\sigma) &\approx \int_{\lambda\rho}^{\infty}\dd\nu\; J_{2s}(\nu)\,u^{(0)}(\sigma,\nu)
\nonumber\\
&\approx \int_{\lambda\rho}^{\infty}\dd\nu\; J_{2s}(\lambda\rho) \, u^{(0)}(\sigma,\lambda\rho)\, e^{\lambda\rho/\xieff} \, e^{-\nu/\xieff}
\nonumber \\
&\approx \int_{\lambda\rho}^{\infty} \dd\nu\; \frac{V_{l-1}(S^{l-1})}{2^{n-1}}\, \rho^{2s+l-1}\; e^{(n-1)\lambda}\;u^{(0)}(\sigma,\lambda\rho)\;e^{\lambda\rho/\xieff}\; e^{-\nu/\xieff}
\nonumber \\
&= \frac{V_{l-1}(S^{l-1})}{2^{n-1}}\, \rho^{l+2s}\; e^{(n-1)\lambda}\; u^{(0)}(\sigma,\lambda\rho)\; \frac{\xieff}{\rho}\,.
\end{align}
We note that
\begin{equation}
\frac{\mu^{(0)}_{2s+2}(\sigma)}{\mu^{(0)}_{2s}(\sigma)} = \rho^{2}= \frac{1}{\lvert k \rvert}\,.
\label{eq:ratio-long}
\end{equation}

The first observation is that the transverse energy
\begin{equation}
\mu^{(0)}_{0}(\sigma) = \left(\frac{V_{l-1}(S^{l-1})}{2^{n-1}}\, \rho^{l}\;e^{(n-1)\lambda}\, u^{(0)}(\sigma,\lambda\rho)\right)\; \frac{\xieff}{\rho}
\end{equation}
consists of two factors. The first factor inside the parentheses is of the Kaluza-Klein type because it is a volume factor $\rho^{l}$ times an energy density. The second factor is novel. It is a potentially large enhancement factor given by the ratio $\xieff/\rho$. This enhancement causes a type of ``critical behavior'' when the energy correlation length $\xi$ approaches $\rho/(n-1)$. Models of this type can  have an enhanced cosmological constant and an enhanced gravitational Planck mass while maintaining the same ratio \eqref{eq:ratio-long} that occurs in Kaluza-Klein compactifications.
	
In the scenarios discussed in this paper where we have tubes embedded in $\AdS_{n}$, we see that $\Lambda_{q} \sim T_{q} G_q \sim \mu_0^{(0)}/\mu_2^{(0)} \sim 1/\rho^{2}= \lvert k \rvert$. Let us recall briefly the results from the same calculation in flat space ($k=0$), see~\cite{alvarez:emergentflat}. In that instance, we have that $\Lambda_{q} \sim T_{q} G_q \sim \mu_0^{(0)}/\mu_2^{(0)} \sim 1/\xi_{\perp}^2$, where $\xi_{\perp}$ is the correlation length for the transverse energy density in flat space. In flat space, $\Lambda_{q}$ is determined by the masses of the particles of the field theory. In both the $\AdS_{n}$ embedding scenarios discussed here, $\Lambda_{q}$ is determined by curvature of $\AdS_{n}$.

A quick back-of-the-envelope calculation shows that, given the current value of the dark energy density $\Omega_\Lambda = 0.685$ and the current scale factor $c^2/3H_0^2 = 6.3\times 10^{51}~\text{m}^{2}$~\cite{pdg:astroconstants}, the cosmological constant is
	\begin{equation}
	\Lambda_{\text{PDG}} = \frac{3 H_0^2 \Omega_\Lambda}{c^2} = 1.1\times 10^{-52}~\text{m}^{-2}
	\end{equation}
The scenarios discussed in this article, applying eq.~\eqref{eq:ratio-short} or~\eqref{eq:ratio-long}, require a radius of curvature $\rho\sim 10^{10}~\text{ly} \sim 10^{26}~\text{m}$ for $\AdS_{4+l}$. This length scale is roughly the size of the observable universe. It is also order of magnitude consistent with the spread allowed by errors in the value of the curvature density $\Omega_{K}= -k/R_{0}^{2}H_{0}^{2}= -0.005^{+0.016}_{-0.017}$ where we used a Hubble length $c/H_{0} \approx 1.4 \times 10^{26}~\text{m}$~\cite{pdg:astroconstants}. These scenarios give reasonable values for cosmological constant $\Lambda_{4}$.
	
Motivated by \cite{ArkaniHamed:1998rs}, we look at the scenario depicted in Figure~\ref{fig:short-range} where $q=4$ and we assume that the energy scale of the $n=4+l$ theory is given by a mass scale $\mu_{n}$ for the $n$-dimensional field theory.	In this case we expect the Kaluza-Klein-like answer for the four-dimensional gravitational constant $(\MPl_{4})^{2} \sim \rho^{l+2} \mu_{n}^{4+l}$. We remark that $\rho\MPl_{4} \sim 10^{61}$. A little algebra leads to 
\begin{equation*}
\frac{\mu_{4+l}}{1~\text{eV}} \sim \frac{\MPl_{4}}{1~\text{eV}} \left(\frac{1}{\rho\MPl_{4}}\right)^{(l+2)/(l+4)} \sim 10^{-33 + 122/(l+4)}
\end{equation*}
This is a very low energy scale with $\mu_{5}\sim 10^{-9}~\text{eV}$, $\mu_{6}\sim 10^{-13}~\text{eV}$. Such energy scales could arise in a conformal field theory in $\AdS_{4+l}$ where the conformal symmetry is softly broken.

We now try to refine this argument. Typically we start with an $n$-dimensional field theory that has a $q$-dimensional world brane representing the defect. In constructing the defect, the field equations are solved using a separation of variables technique assuming  a Cartesian decomposition of the spacetime of the form $\Sigma^{q}\times \mathbb{E}^{l}$. The field configuration is determined by solving the field theory in the $l$-dimensional transverse space. Thus, it is natural to assume that the mass scale for the full Lagrangian of the $n$-dimensional field theory, rather than a simple $\mu_{n}^{n}$,  should have the product form $\mu_{\parallel}^{q}\mu_{\perp}^{l}$ where $\mu_{\perp}$ is a mass scale for the transverse $l$-dimensional field theory  and $\mu_{\parallel}$ is an overall scale originating from the details of the full $n$-dimensional Lagrangian. The scenario depicted in Figure~\ref{fig:short-range} leads to the relation $(\MPl_{q})^{q-2} = \rho^{l+2} \mu_{\parallel}^{q}\mu_{\perp}^{l}$, or
\begin{equation}
\rho\mu_{\parallel} = \frac{\left(\rho\MPl_{q}\right)^{1-2/q}}{\left(\rho\mu_{\perp}\right)^{l/q}}
\end{equation}
Specializing to $q=4$ we obtain
\begin{equation}
\rho\mu_{\parallel} = \frac{\left(\rho\MPl_{4}\right)^{1/2}}{\left(\rho\mu_{\perp}\right)^{l/4}}
\end{equation}
The cosmological parameters give $\rho\MPl_{4} \sim 10^{61}$, and in our scenario we require $\rho \ll 1/\mu_{\perp}$. We obtain
\begin{equation}
\rho\mu_{\parallel} = 10^{30.5 + (l/4)\log_{10}(1/\rho\mu_{\perp})}
\end{equation}
\begin{equation}
\frac{\mu_{\parallel}}{1~\text{eV}} = 10^{-2.5 + (l/4)\log_{10}(1/\rho\mu_{\perp})}
\end{equation}
For the sake of computational simplicity, we assume that $\rho\mu_{\perp} \approx 10^{-2}$; this corresponds to $\mu_{\perp}\sim 10^{-35}~\text{eV}$. Thus we find that $\mu_{\parallel}\sim 10^{-2.5 + l/2}~\text{eV}$. An energy scale of $1~\text{eV}$ corresponds to a length scale of $10^{-6}~\text{m}$.

In the second scenario, see Figure~\ref{fig:long-range}, we have a similar relationship but with  an enhancement factor  of $\xieff/\rho$:
\begin{equation}
(\MPl_{q})^{q-2} = \rho^{l+2} \mu_{\parallel}^{q}\mu_{\perp}^{l}\cdot\frac{\xieff}{\rho}
\end{equation}
The enhancement factor could be quite large and may serve to decrease the product $\mu_{\parallel}^{q}\mu_{\perp}^{l}$.
We find that the longitudinal scale is given by
\begin{equation}
\rho\mu_{\parallel} = \frac{\left(\rho\MPl_{q}\right)^{1-2/q}}{\left(\rho\mu_{\perp}\right)^{l/q}} \left( \frac{\xieff}{\rho}\right)^{-1/q}
\end{equation}
If we specialize to $q=4$, the above becomes
\begin{equation}
\rho\mu_{\parallel} = \frac{\left(\rho\MPl_{q}\right)^{1/2}}{\left(\rho\mu_{\perp}\right)^{l/4}} \left( \frac{\xieff}{\rho}\right)^{-1/4}
\end{equation}
Putting in the same numbers as before we have
\begin{equation}
\mu_{\parallel} = \left( \frac{\xieff}{\rho}\right)^{-1/4} \cdot 10^{-2.5 + (l/4)\log_{10}(1/\rho\mu_{\perp})}\;\text{ eV}
\end{equation}
	
\section{Conclusions}
	

In this article we discussed a mechanism where non-gravitational physics in higher dimension induces an emergent theory of gravity. Our premise begins with the assumption that we have a $q$-submanifold $\Sigma^{q}$ embedded in $\AdS_{n}$ and that the energy (action) of our model is localized in a tubular neighborhood of $\Sigma^{q}$. The detailed reason for this localization is left unexplained but we assume it arises from an underlying higher-dimensional field theory without gravity.  We derived a general framework that leads to an effective Lagrangian that describes the dynamics of $\Sigma^{q}$. This effective Lagrangian is a  Lovelock gravitational theory and the multipole moment coefficients are the coupling parameters of the Lagrangian. 

In the more traditional brane scenarios of AHDD, RS, and DGP, the four-dimensional gravitational constant is in effect a consequence of higher-dimensional gravity with an appropriately chosen gravitational mass scale $\MPl_{n}$ and an effective length scale that arises differently in the various  brane scenarios related to the relationship between the embedded submanifold $\Sigma^{q}$ and the ambient $n$-manifold $M^{n}$. In AHDD, the length scale is the size of the Kaluza-Klein compactification manifold that for them is in the millimeter scale because they choose the higher dimensional gravitational scale to be in the TeV range. In the RS scenario, their $\AdS_{5}$ has a radius of curvature that is on the order of the Planck length while their ``compactification scale'' $r_{c}$ is a couple of orders of magnitude larger. An interesting consequence of RS is that the four-dimensional Planck scale $\MPl_{4}$ is insensitive to their compactification scale and depends on the $5$-dimensional gravity scale $\MPl_{5}$ and the radius of curvature of the $\AdS_{5}$. Our scenario produces a different result but similar in spirit. The DGP mechanism has  $5$-dimensional gravity with a TeV scale, and  a crossover scale from higher-dimensional to lower-dimensional gravity on the order of the size of the solar system. However, the scenario that DGP use involves extra-space dimensions that are flat and infinitely sized; there is no underlying curvature in the bulk. 

In our model there is no higher dimensional gravitation: gravity emerges from non-gravitational physics and is not induced from higher dimensional gravity. The higher-dimensional energy scale is set by the higher-dimensional non-gravitational field theory and not by a higher-dimensional gravitational constant. We have a mechanism in which the effective compactification radius arises naturally from the radius of curvature of spacetime which we take to be compatible with the experimental bounds. 
In  our model, the energy localization is over cosmological distances comparable to the radius of curvature of the $\AdS_{n}$. The first result is that the $q$-dimensional cosmological constant $\Lambda_{q}\sim 1/\rho^{2}= \lvert k \rvert$. A second result is that we can reproduce $\MPl_{4}\sim 10^{19}~\text{GeV}$ if we assume that the energy scale of the progenitor field theory\footnote{The effective $n$-dimensional energy scale $\mu$ is defined by $\mu^{4+l} = \mu_{\parallel}^{4}\mu_{\perp}^{l}$, see Section~\ref{sec:energy-action}.} in $\AdS_{n}$ is very low $O(10^{-22})~\text{GeV}$. One can imagine beginning with a conformal field theory in $\AdS_{n}$ that, through some type of conformal symmetry breaking, leads to a minuscule energy scale determined by the underlying radius of curvature $\rho$ of the $\AdS_{n}$. Because there is no higher dimensional gravity, there is no crossover behavior from higher dimensional gravity to gravity on the brane, and we avoid this issue entirely.

	\acknowledgments

	This work was supported in part by the
	National Science Foundation under Grant PHY-1212337.

	\bibliographystyle{jhep}
	\providecommand{\href}[2]{#2}\begingroup\raggedright\endgroup

\end{document}